\documentclass{elsart}
\usepackage{graphicx,amssymb}
\journal{Physica A}

\begin{document}
\begin{frontmatter}
\title{Random Matrix Theory and Fund of Funds Portfolio Optimisation}
\author[DCU]{T. Conlon},
\ead{tconlon@computing.dcu.ie}
\author[DCU]{H.J. Ruskin},
\ead{hruskin@computing.dcu.ie}
\author[DCU]{M. Crane},
\ead{mcrane@computing.dcu.ie}
\address[DCU]{Dublin City University, Glasnevin, Dublin 9, Ireland}

\begin{abstract}
The proprietary nature of Hedge Fund investing means that it is common practise for managers to release minimal information about their returns.  The construction of a Fund of Hedge Funds portfolio requires a correlation matrix which often has to be estimated using a relatively small sample of monthly returns data which induces noise.  In this paper random matrix theory (RMT) is applied to a cross-correlation matrix C, constructed using hedge fund returns data.  The analysis reveals a number of eigenvalues that deviate from the spectrum suggested by RMT.  The components of the deviating eigenvectors are found to correspond to distinct \emph{groups of strategies} that are applied by hedge fund managers.  The Inverse Participation ratio is used to quantify the number of components that participate in each eigenvector.  Finally, the correlation matrix is cleaned by separating the noisy part from the non-noisy part of C.  This technique is found to greatly reduce the difference between the predicted and realised risk of a portfolio, leading to an improved risk profile for a fund of hedge funds.
\end{abstract}
\begin{keyword}
Random Matrix Theory \sep Hedge Funds \sep Fund of Funds \sep Correlation Matrix \sep Portfolio Optimisation. 
\end{keyword}
\end{frontmatter}

\section{Introduction}
A Hedge Fund is a lightly-regulated private investment vehicle that may utilise a wide range of investment strategies and instruments. These funds may use short positions, derivatives, leverage and charge incentive-based fees. Normally, they are structured as limited partnerships or offshore investment companies.  Hedge Funds pursue positive returns in all markets and hence are described as ``absolute return'' strategies. 

Hedge Funds are utilised by pension funds, high net-worth individuals and institutions, due to their low correlation to traditional long-only investment strategies.  The incentive-based performance fees, earned by hedge fund managers, align the interest of the hedge fund manager with that of the investor.  The performance of Hedge Funds has been impressive, with the various Hedge Fund indices providing higher returns, with lower volatility, than traditional assets over many years.  As of the end of the first quarter 2006 the total assets managed by hedge funds world wide is estimated at \$1.25 trillion \cite{Barclays}.  Hedge Funds generally only report their returns on a monthly basis and this means that there is a very limited amount of data available to study as databases of Hedge Fund returns have only been in operation for about 15 years.  This is in keeping with the highly secretive, proprietary nature of Hedge Fund investing.  The amount of information reported by a Hedge Fund about how and where it is producing its returns is often limited to sectoral overviews and strategy allocations.  For an introduction to hedge funds see \cite{Hedge_Funds_intro,Hedge_Funds_quant}.

Significant diversification benefits can be gained by investing in a variety of hedge fund strategies, due to the presence of low and even negative correlations between different hedge fund strategies.  Such strategies can be broken up into two general categories: directional and market neutral.  Directional strategies, (for example Long/Short Equity, Emerging markets, Macro and Managed Futures) have a high risk, high return profile and act as return enhancers to a traditional portfolio.  Market neutral strategies, (for example Convertible Arbitrage, Equity Market Neutral and Fixed Income Arbitrage) deploy a low risk profile and act as a substitute for some proportion of the fixed income holdings in an investors portfolio, \cite{Hedge_Funds_intro,Hedge_Funds_quant}. 

A Fund of Hedge Funds allows investors to have access to a large diverse portfolio of Hedge Funds without having to carry out due diligence on each individual manager.  The diversification benefits provided by Fund of Funds are brought about by investing in a number of funds that have a low correlation to each other.  These correlations are often calculated by using equally weighted fund returns and can contain a significant amount of noise due to the very small amount of returns data available for hedge funds \cite{Hedge_Funds_quant}. 

In this paper we apply Random Matrix Theory to Hedge Fund returns data with the aim of reducing the levels of noise in these correlation matrices formed from this data and hence constructing a fund of hedge funds with an improved risk profile.  Previous studies have used the information found in the RMT defined deviating eigenvalues of a correlation matrix as inputs into a minimum spanning tree \cite{Miceli_2004} to enable characterisation of Hedge Fund strategies.  In this paper the components of the deviating eigenvectors are shown to correspond to distinct groups of strategies that are applied by hedge fund managers and this is exploited to construct a portfolio with reduced levels of risk.

This paper is organised as follows:  in Section 2 we review RMT and discuss its use in the extraction of information from a correlation matrix of Hedge Fund returns using RMT techniques.  In Section 3 we look at the results obtained applying Random Matrix theory to Hedge Funds and, in the final section, we draw our conclusions.

\section{Methods}
\subsection{Random Matrix Theory}
Given returns $G_{i} \left(t\right)$, $i = 1, \ldots,N$, of a collection of Hedge Funds we define a normalised return in order to standardise the different fund volatilities. We normalise $G_{i}$ with respect to its standard deviation $\sigma_{i}$ as follows:
\begin{equation}
g_{i}\left(t\right) = \frac{G_{i} \left(t\right) - \widehat{G_{i} \left(t\right)}}{\sigma_{i}}
\end{equation}
Where $\sigma_{i}$ is the standard deviation of $G_{i}$ for assets $i = 1, \ldots,N$ and $\widehat{G_{i}}$ is the time average of $G_{i}$ over the period studied.

Then the equal time cross correlation matrix is expressed in terms of $g_{i}\left(t\right)$
\begin{equation}
C_{ij} \equiv \left\langle g_{i}\left(t\right) g_{j}\left(t\right) \right\rangle.
\end{equation}
The elements of $C_{ij}$ are limited to the domain $-1 \leq C_{ij} \leq 1$, where $C_{ij} = 1$ defines perfect correlation between funds, $C_{ij} = -1$ corresponds to perfect anti-correlation and $C_{ij} = 0$ corresponds to uncorrelated funds. In matrix notation, the correlation matrix can be expressed as
\begin{equation}
\mathbf{C} = \frac{1}{T} \mathbf{GG}^{t}
\end{equation}
Where $\bf{G}$ is an $N\times T$ matrix with elements $g_{it}$.

The spectral properties of $\bf{C}$ may be compared  to those of a ``random'' Wishart correlation matrix \cite{Laloux_2000,Plerou_2002},
\begin{equation}
\mathbf{R} = \frac{1}{T} {\mathbf{AA}}^{t}
\end{equation}
Where $\bf{A}$ is an $N\times T$ matrix with each element random with zero mean and unit variance.
Statistical properties of random matrices have been known for many years in the physics literature \cite{Mehta_2004} and have been applied to financial problems relatively recently \cite{Bouchaud_book,Laloux_1999,Laloux_2000,Plerou_1999,Gopikrishnan_2000,Plerou_2002,Rosenow_2002,Burda_2004_a,Burda_2004_b,Wilcox_2004,Sharifi_2004}.  

In particular, the limiting property for the sample size $N \rightarrow \infty$ and sample length $T \rightarrow \infty$, providing that $Q = \frac{T}{N} \geq 1$ is fixed, has been examined to show analytically that the distribution of eigenvalues $\lambda$ of the random correlation matrix $\bf{R}$ is given by:
\begin{equation}
P_{rm} \left(\lambda\right) = \frac{Q}{2\pi \sigma^{2}} \frac{\sqrt{\left(\lambda_{+} - \lambda\right)\left(\lambda - \lambda_{-}\right)}}{\lambda}  ,
\label{Theo_evalue_dist}
\end{equation}
for $\lambda$ within the region $\lambda_{-} \leq \lambda_{i} \leq \lambda_{+}$, where $\lambda_{-}$ and $\lambda_{+}$ are given by
\begin{equation}
\lambda_{\pm} = \sigma^{2}\left(1 + \frac{1}{Q} \pm 2\sqrt{\frac{1}{Q}}\right),
\end{equation}
Where $\sigma^{2}$ is the variance of the elements of $\bf{G}$; (for $\bf{G}$ normalised this is equal to unity).

$\lambda_{\pm}$ are the bounds of the theoretical eigenvalue distribution.  Eigenvalues that are outside this region are said to deviate from Random Matrix Theory.  Hence by comparing the empirical distribution of the eigenvalues of the funds correlation matrix to the distribution for a random matrix as given in Equation ~\ref{Theo_evalue_dist}, we can identify the deviating eigenvalues.  These deviating eigenvalues are said to contain information about the system under consideration and we use eigenvector analysis to identify specifically the information present.

\subsection{Eigenvector Analysis}
\label{Eig_anal}
The Deviations of $P(\lambda)$ from the RMT result $P_{rm}(\lambda)$ implies that these deviations should also be displayed in the statistics of the corresponding eigenvector components \cite{Laloux_2000}. In order to interpret the meaning of the deviating eigenvectors, we note that the largest eigenvalue is of an order of magnitude larger than the others, which constrains the remaining $N-1$ eigenvalues since $Tr \left[C\right] = N$. Thus, in order to analyse the contents of the remaining eigenvectors, we first remove the effect of the largest eigenvalue.  To do this we use the linear regression (as discussed in \cite{Plerou_2002})
\begin{equation}
G_{i}\left(t\right) = \alpha_{i} + \beta_{i}G^{large}\left(t\right) + \epsilon_{i}\left(t\right),
\end{equation}
Where $G^{large} = \sum^{N}_{1} u^{large}_{i}G_{i}\left(t\right)$ and $N = 49$ is the number of funds in our sample.  Here $u^{large}_{i}$ corresponds to the components of the largest eigenvector.  We then recalculate the correlation matrix C using the residuals $\epsilon_{i}\left(t\right)$.  If we quantify the variance of the part not explained by the largest eigenvalue as $\sigma^{2} = 1 - \lambda_{large}/n$, \cite{Laloux_2000}, we can use this value to recalculate our values of $\lambda_{\pm}$.

Using techniques for sector identification, \cite{Gopikrishnan_2000}, we try to analyse the information contained in the eigenvectors.  We partition the funds into groups $l = 1, \ldots 10$ as defined by the managers strategy (eg. Equity Long/Short, Managed Futures. See Appendix A for a complete strategy breakdown for the sample). We define a projection matrix $P_{li} = \frac{1}{n_{l}}$, if fund $i$ belongs to group $l$ and $P_{li} = 0$ otherwise. For each deviating eigenvector $u^{k}$ we compute the contribution $X^{k}_{l} \equiv \sum^{N}_{i=1} P^{k}_{li} \left[ u^{k}_{i} \right]^{2}$ of each strategy group, where this represents the product of the projection matrix and the square of the eigenvector components.  
This allows us to measure the contribution of the different Hedge Fund strategies to each of the eigenvectors.

As suggested in \cite{Plerou_2002}, we also aim to assess how the effects of randomness lessen as we move further from the RMT upper boundary boundary, $\lambda_{+}$.  To do this we use the Inverse Participation Ratio (IPR).  The IPR allows quantification of the number of components that participate significantly in each eigenvector and tells us more about the level and nature of deviation from RMT. The IPR of the eigenvector $u^{k}$ is given by
$ I^{k} \equiv \sum^{N}_{l = 1} \left( u^{k}_{l}\right)^{4} $ and allows us to compute the inverse of the number of eigenvector components that contribute significantly to each eigenvector.

\subsection{Application to portfolio optimisation}
\label{Port_Opt_Intro}
The diversification of an investment into independently fluctuating assets reduces its risk.  However, since cross-correlations between asset prices exist, the accurate calculation of the cross-correlation matrix is vitally important.  The return on a portfolio with $N$ assets is given by $\Phi = \sum^{N}_{i=1} w_{i} G_{i}$ where $G_{i}(t)$ is the return on asset $i$, $w_{i}$ is the fraction of wealth invested in asset $i$ and $\sum^{N}_{i=1} w_{i} =1$.  The risk of holding this portfolio is then given by
\begin{equation}
\Omega^{2} = \sum^{N}_{i=1} \sum^{N}_{j=1} w_{i} w_{j} C_{ij} \sigma_{i} \sigma_{j}
\end{equation}
Where $\sigma_{i}$ is the volatility of $G_{i}$ and $C_{ij}$ are the elements of the cross-correlation matrix.  In order to find an optimal portfolio, using the Markowitz theory of portfolio optimisation \cite{Bouchaud_book,Markowitz_1959,Elton_Gruber_2002}, we minimise $\Omega^{2}$ under the constraint that the return on the portfolio, $\Phi$, is some fixed value.  This minimisation can be implemented by using two Lagrange multipliers, which leads to a set of $N$ linear equations which can be solved for $w_{i}$. If we minimise $\Omega$ for a number of different values of $\Phi$, then we can obtain a region bounded by an upward-sloping curve, called the \emph{efficient frontier}.

In the case of optimisation with a portfolio containing only Hedge Funds, the additional constraint of no short-selling is natural due to the difficulties in short-selling of funds; (note that short-selling may be achievable by the use of swaps but is uncommon) \cite{Hedge_Funds_intro,Hedge_Funds_quant}.

In order to demonstrate the effects of randomness on the cross-correlation matrix of hedge funds we first divide the time series studied into two equal parts.  We assume that we have \emph{perfect knowledge} on the future average returns $m_{i}$ by taking the observed returns on the second sub-period.  We calculate
\begin{enumerate}
\item
the predicted efficient frontier using the correlation matrix for the first sub-period and the expected returns $m_{i}$
\item
the realised efficient frontier using the correlation matrix for the second sub-period and the expected returns $m_{i}$.
\end{enumerate}
The portfolio risk due to the noise in the correlation matrix can then be calculated using
\begin{equation}
R_{p} = \frac{\Omega_{r}^2 - \Omega_{p}^2}{\Omega_{p}^2}  
\end{equation}
Where $\Omega_{r}^2$ is the risk of the realised portfolio and $\Omega_{p}^2$ is the risk of the predicted portfolio.

It was shown in \cite{Burda_2004_a, Burda_2004_b} that correlations may also be measured in the random part of the eigenvalue spectrum.  However, since our aim here is to demonstrate how Random Matrix Theory can be used to improve the risk/return profile for a portfolio of Hedge Funds, we assume that the eigenvalues corresponding to the noise band in RMT, $\lambda_{-} \leq \lambda \leq \lambda_{+}$, are not expected to correspond to real information following \cite{Bouchaud_book,Laloux_1999,Laloux_2000,Plerou_1999,Gopikrishnan_2000,Plerou_2002,Rosenow_2002,Wilcox_2004,Sharifi_2004}. We then use this assumption to remove some of the noise from the correlation matrix.  Although the technique used in \cite{Bouchaud_book,Laloux_2000} has been shown to lead to problems with the \emph{stability} of the correlation matrix \cite{Sharifi_2004}, we apply it here as a simple test case to demonstrate how noise can be removed from the cross-correlation matrix formed from hedge fund returns.  This technique involves separating the noisy and non-noisy eigenvalues and keeping the non-noisy eigenvalues the same.  The noisy eigenvalues are then replaced by their average and the correlation matrix is reconstructed from the cleaned eigenvalues.  We can then compare the risk of both the predicted and realised portfolios for the original and cleaned correlation matrices.

\newpage

\section{Results}
\subsection{Equally weighted Correlation Matrix}
The dataset studied here is a collection of 49 Hedge Funds with varying strategies over a synchronous period from January 1997 to September 2005 $\left(T = 105\right)$.   The original dataset was much larger (approximately 1500 funds) but since the length of data available was much less than the number of funds we were forced to choose a subset. The subset chosen were the 49 funds with the longest track records giving us a fund to data ratio $Q = 2.143$. Reducing the dataset in this way is not unrealistic as a typical fund of hedge funds would monitor a subset of funds and choose a portfolio from these \cite{Hedge_Funds_intro,Hedge_Funds_quant}. Often one of the criteria used in choosing this subset of investable funds would be the completion of a minimum track record. Other data sets, such as a portfolio made up of Hedge Fund strategy indices, were also studied by the authors with notably similar results.

\begin{figure}[htbp!]
\begin{center}
\includegraphics[height=77mm,width=110mm]{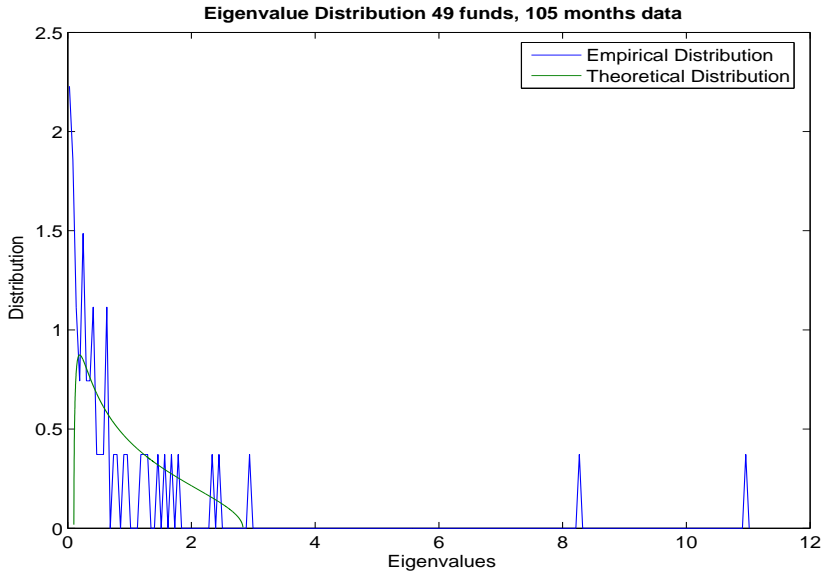}
\caption{Spectral Distribution for equally weighted Hedge Fund correlation matrix}
\label{evalue_dist}
\end{center}
\end{figure}

First we calculated the empirical Correlation Matrix between the funds using equally weighted returns and from this found the spectrum of eigenvalues.  This is then compared with the theoretical spectrum for random Wishart matrices (as per equation ~\ref{Theo_evalue_dist}), using $\lambda_{-} = 0.1$ and $\lambda_{+} = 2.83$.  As can be seen from Figure~\ref{evalue_dist}, the bulk of the eigenvalues conform to those of the random matrix.  There are three deviating eigenvalues, at $10.9886$, $8.2898$ and $2.944$. This means that $6.1\%$ of eigenvalues deviate from the RMT prediction which is consistent with the findings of \cite{Laloux_2000}, where the authors argue that at most $6 \%$ of eigenvalues are non-noisy.

\newpage

\subsection{Bootstrapping}
\begin{figure}[htbp!]
\begin{center}
\includegraphics[height=110mm,width=110mm]{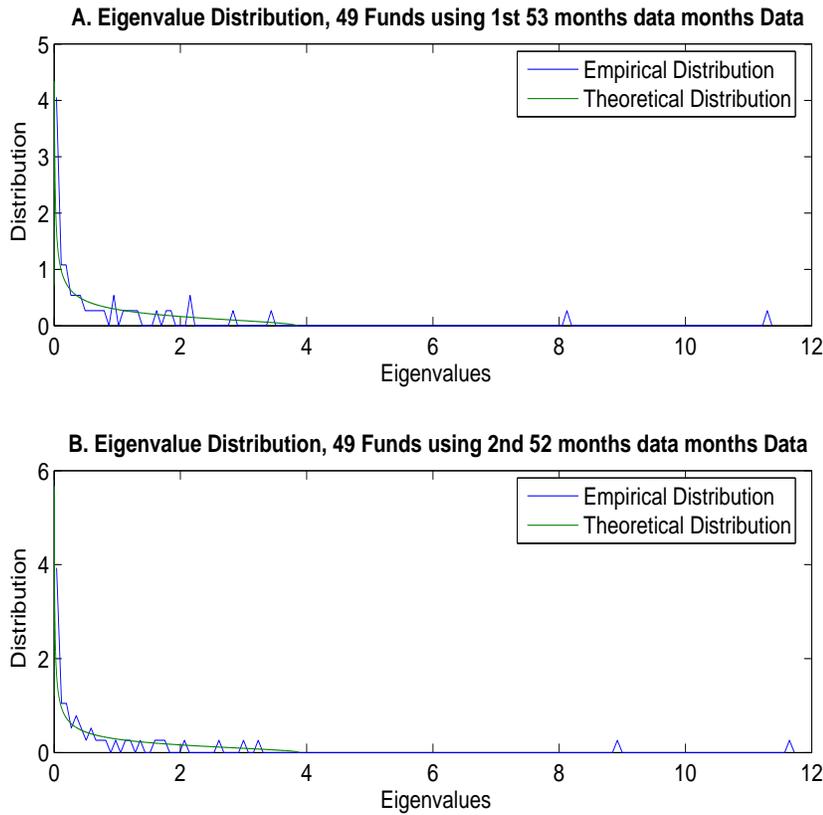}
\caption{Bootstrapped spectral distribution for consecutive periods.}
\label{bootstrapped_results}
\end{center}
\end{figure}

In order to show that there was no dependence on the choice of time period or the length of the series we broke the time series up into two segments and again compared the eigenvalue spectrum of the correlation matrix with that of a random Wishart matrix.  For both time periods studied we found just two eigenvalues that deviated from the RMT prediction. As can be seen in Figure~\ref{bootstrapped_results}, the anomalous eigenvalue contributions are very similar for both periods chosen, which implies independence from the choice of time period and stationarity of the data.  The values of the deviating eigenvalues are shown in table 1.

\begin{table}[htbp!]
	\centering
		\begin{tabular}{c||c|c|c}
		\emph{Eigenvalue Rank} & \emph{105 Months Returns} &  \emph{$1^{st}$ 53 Months}  & \emph{$2^{nd}$ 52 Months}\\
		\hline
		\hline
			1 & 10.9886	&		11.334	&	11.6874\\
			2 & 8.2898	&		8.1375	&	8.9312\\
			3 & 2.944		&	  	&	
		\end{tabular}
	\caption{Deviating Eigenvalues}
	\label{tab:Eigenvalues}
\end{table}

\newpage

\subsection{Eigenvector Analysis}
\label{EigVec_anal}
\begin{figure}[htbp!]
\begin{center}
\includegraphics[height=65mm,width=120mm]{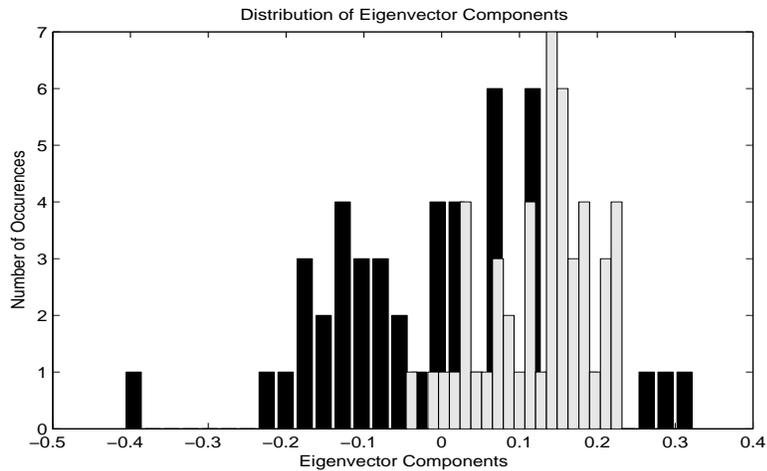}
\caption{Comparison of Eigenvector Components, largest Eigenvector (Grey), Eigenvector from the bulk (Black)}
\label{Eigenvector_Components}
\end{center}
\end{figure}

Figure~\ref{Eigenvector_Components} shows the distribution of the components of the largest eigenvector and also the components of a typical eigenvector from the region predicted by RMT.  As can be seen from this graph, the distribution of the components of the largest eigenvector are significantly different from that of an eigenvector chosen from the random region.  The average value is much larger and the variance of the components much smaller than for the largest eigenvector, which is in agreement with \cite{Plerou_2002,Sharifi_2004} (the largest eigenvector can be interpreted as the `market').  In this case the `market' is the set of external stimuli that affect most hedge funds, (eg. Interest rate changes, large market (ie S$\&$P 500 etc) moves, margin changes etc).

\begin{figure}[htbp!]
\begin{center}
\includegraphics[height=70mm,width=110mm]{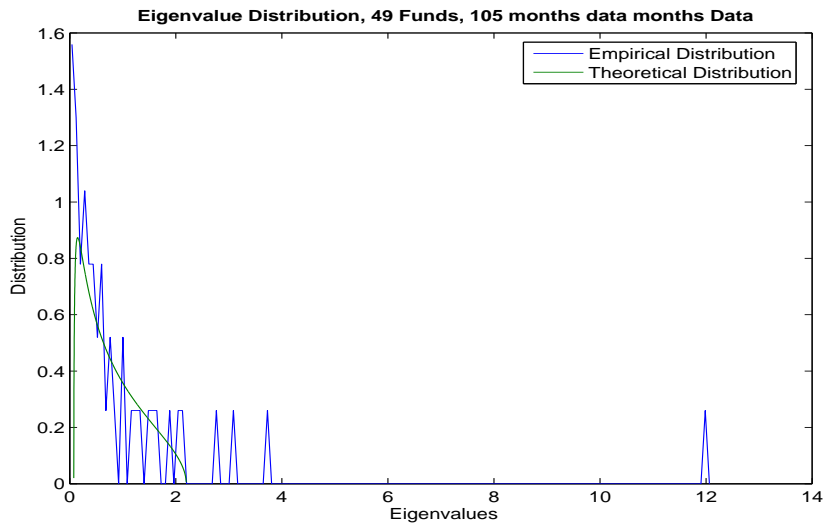}
\caption{Eigenvalue spectrum after the removal of the effects of the largest Eigenvalue}
\label{Eigenvalue_Spectrum}
\end{center}
\end{figure}

We then remove the effects of the largest eigenvalue using the techniques described in Section \ref{Eig_anal}.  This changes the value of $\lambda_{max}$ from $2.8329$ to $2.1975$ which means that 4 of the remaining largest eigenvalues are now outside the RMT region, (Figure~\ref{Eigenvalue_Spectrum}).  

\begin{figure}[htbp!]
\begin{center}
\includegraphics[height=70mm,width=120mm]{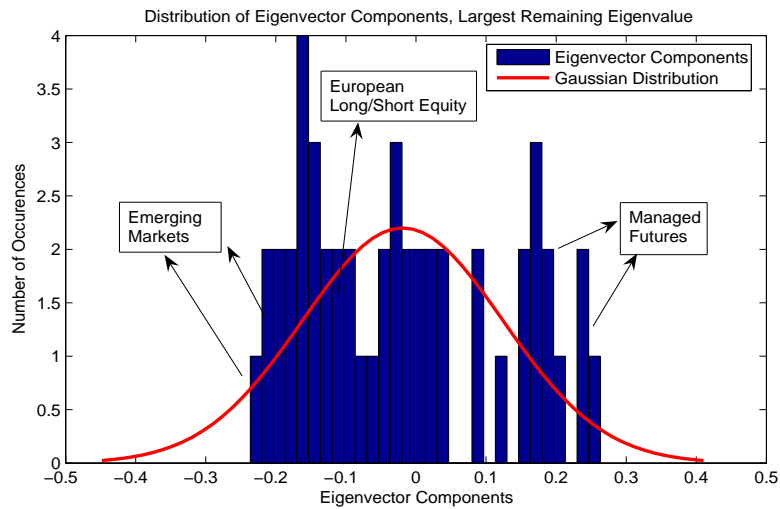}
\caption{Distribution of eigenvector components, largest remaining eigenvalue}
\label{largest_eigenvector}
\end{center}
\end{figure}

The distribution of the components of largest remaining deviating eigenvector shows some distinctive clustering (Figure~\ref{largest_eigenvector}).  In particular the Managed Futures, Emerging markets and European long/short equity strategies are the major contributors here. 

\begin{figure}[htbp!]
\begin{center}
\includegraphics[height=70mm,width=120mm]{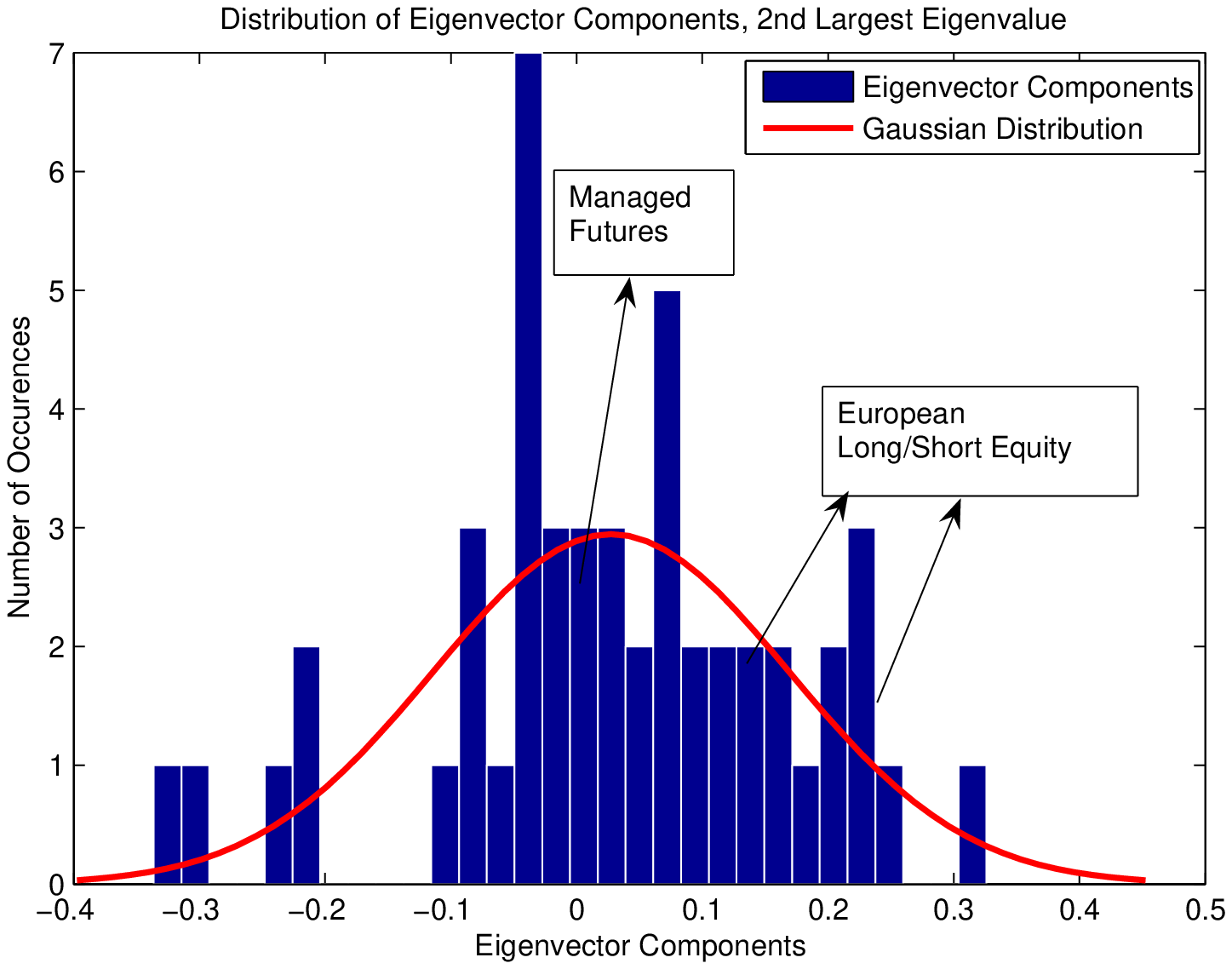}
\caption{Distribution of eigenvector components, 2nd largest remaining eigenvalue}
\label{2nd_largest_eigenvector}
\end{center}
\end{figure}

A similarly-clustered distribution also emerges for the other deviating eigenvalues.  An analysis of the eigenvector components for the 2nd largest remaining eigenvector, after removing the effects of the market eigenvector (Figure~\ref{2nd_largest_eigenvector}), shows distinctive clusters emerging, especially for the managed futures and long/short equity sectors.  However, the components corresponding to the managed futures strategy do not deviate much from zero and hence make little contribution.  These findings for clustering of the deviating eigenvalues agree with \cite{Sharkasi_2006} where the authors show that, in addition to the largest, the other deviating eigenvalues of the correlation matrix of asset returns also contain information about the risk associated with the assets.

\subsection{Strategy Identification}
\begin{figure}[htbp!]
\begin{center}
\includegraphics[height=65.5mm,width=120mm]{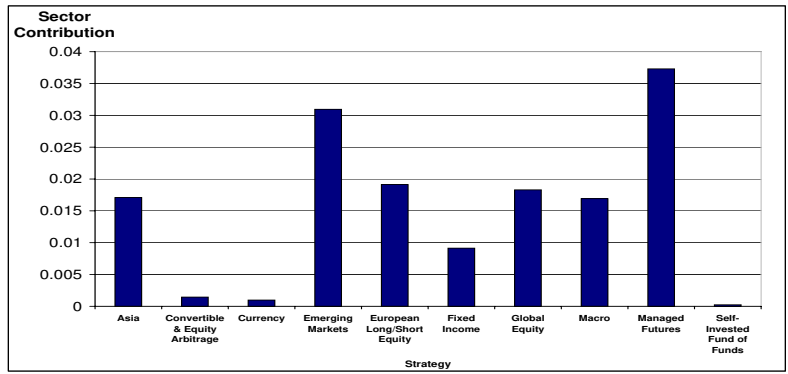}
\caption{Strategy Contribution, largest eigenvalue}
\label{Sector_contrib_2nd_larg_evec}
\end{center}
\end{figure}

From the analysis described in Section \ref{EigVec_anal}, we look at the contribution, $X^{k}_{l}\equiv \sum^{N}_{i=1} P^{k}_{li} \left[ u^{k}_{i} \right]^{2}$, to each of the deviating eigenvalues from the different strategies employed. Figure~\ref{Sector_contrib_2nd_larg_evec} shows $X^{k}_{l}$ for the largest remaining Eigenvector once the effects of the market eigenvalue are removed.  The largest contributors are clearly Managed Futures and Emerging markets. However, the strategy contribution for Managed Futures is only around twice that for many of the other sectors.  Hence Managed Futures and Emerging markets are the dominant strategies but care in interpretation is needed, since neither contributor is dominant overall.

\begin{figure}[htbp!]
\begin{center}
\includegraphics[height=65.5mm,width=120mm]{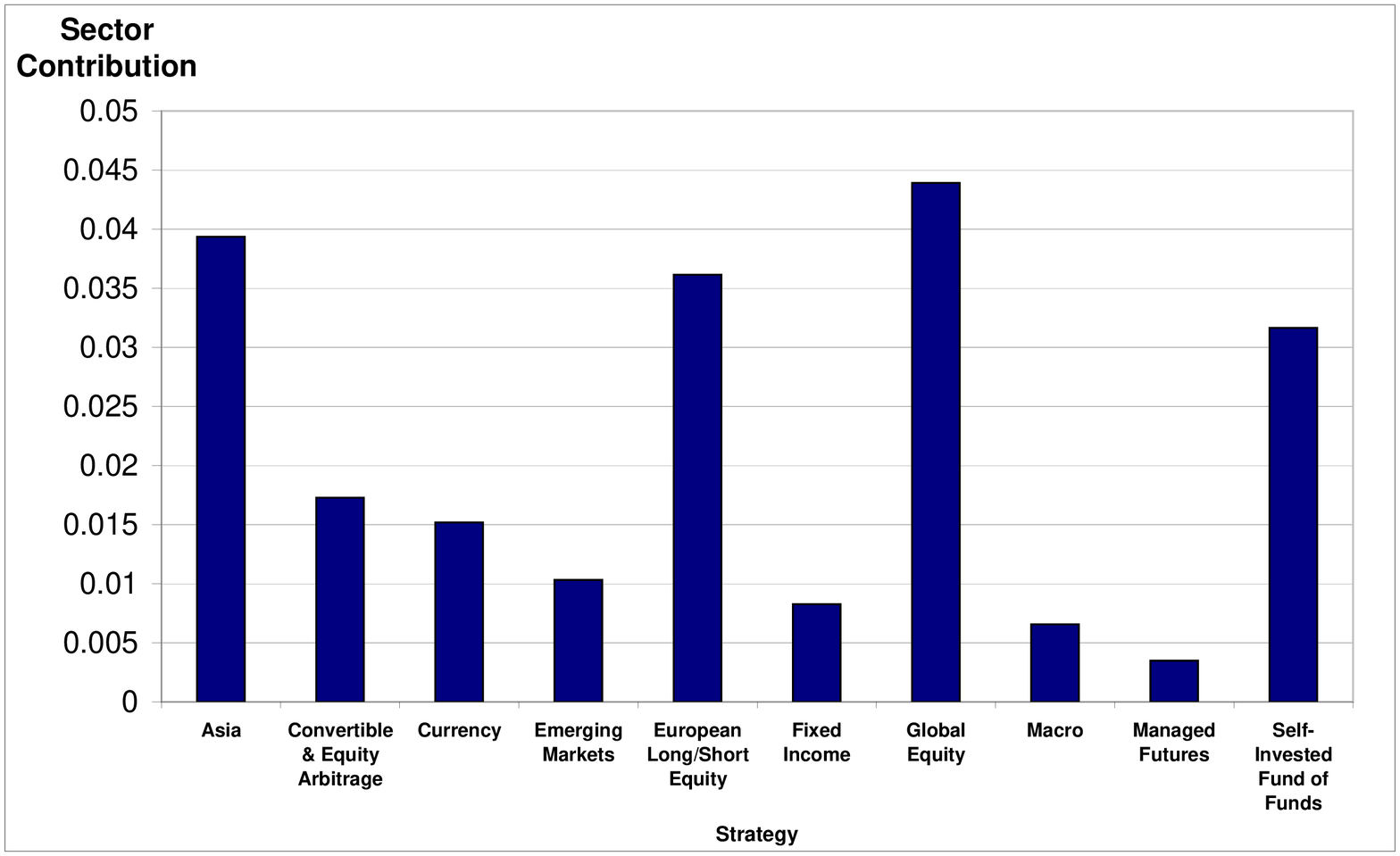}
\caption{Strategy Contribution, 2nd largest eigenvalue}
\label{Sector_contrib_3rd_larg_evec}
\end{center}
\end{figure}

Figure~\ref{Sector_contrib_3rd_larg_evec} shows the strategy contributions for the second largest eigenvalue.  These are interesting, since three of the four dominant strategies (Asia, Global Equity $\&$ European Long/Short Equity) are $\it{equity}$ $\it{strategies}$ and would all be affected by events in world equity markets. Also the fourth strategy, Self Invested Fund of Funds, may well also consist of equity funds. However there is limited information available on exactly what type of funds the managers were invested, although there is reason to believe that a majority of them would be equity based.  This implies that this eigenvalue seems to contain information just on equity funds.

\begin{figure}[htbp!]
\begin{center}
\includegraphics[height=65.5mm,width=120mm]{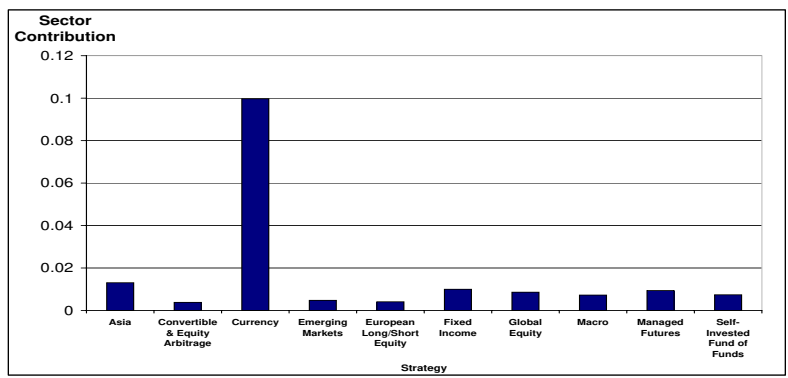}
\caption{Strategy Contribution, 3rd largest eigenvalue}
\label{Sector_contrib_4th_larg_evec}
\end{center}
\end{figure}

Figure~\ref{Sector_contrib_4th_larg_evec} contains the strategy contributions for the third largest remaining eigenvalue.  Clearly the dominant strategy here is Currency.  The final deviating eigenvalue also has one dominant strategy which is Self Invested Fund of Funds.  It is notable in the above that analysis of the eigenvalues from within the random matrix region revealed no dominant strategies.  The evidence of strategy information in the deviating eigenvalues, coupled with a lack of dominant strategies within the RMT region, supports the idea that information in the correlation matrix is cheifly contained within the deviating eigenvalues.  We show how this information can be used to create a portfolio with an improved risk-return profile in Section \ref{Port_Opt}. 

\clearpage

\subsection{Inverse Participation Ratio}
\begin{figure}[htbp!]
\begin{center}
\includegraphics[height=90mm,width=120mm]{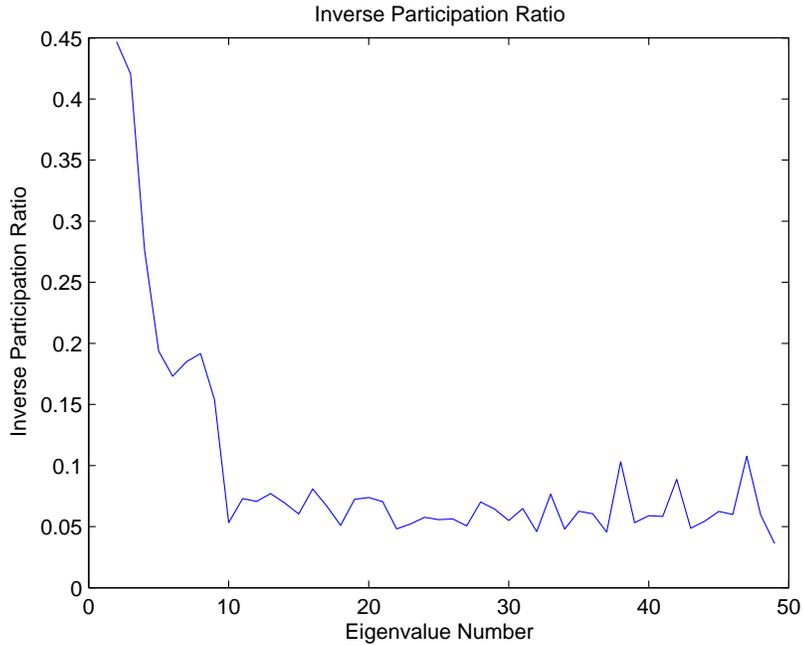}
\caption{Inverse Participation Ratio}
\label{Inverse_Part_Ratio}
\end{center}
\end{figure}

Figure~\ref{Inverse_Part_Ratio} shows the inverse participation ratio (IPR) calculated for the eigenvectors of the Hedge Fund cross-correlation matrix studied. The average IPR value is around 0.06, larger than would be reasonably expected ($\frac{1}{N} \approx 0.02$) if all components contributed to each eigenvector. We would also expect that the largest eigenvectors contributed much more markedly. However they appear to have a similar IPR to eigenvectors corresponding to the random section.  Part of the reason may be that the sample size is small, so the IPR is not particularly effective in terms of assessing by how much larger eigenvectors deviate from the random region, since the average value of the IPR relies on a sample size that tends to infinity. 

A significant deviation from the average IPR value is found for the first few eigenvalues. This seems to have been caused by the inclusion of four or five funds, identical apart from being either in different base currencies or leveraged versions of each other with correspondingly high correlation ($\approx1$). As mentioned in \cite{Plerou_2002} funds with a correlation coefficient much greater than the average effectively decouple from the other funds.

\newpage

\subsection{Noise Removal and Portfolio Optimisation}
\label{Port_Opt}
\begin{figure}[htbp!]
\begin{center}
\includegraphics[height=90mm,width=110mm]{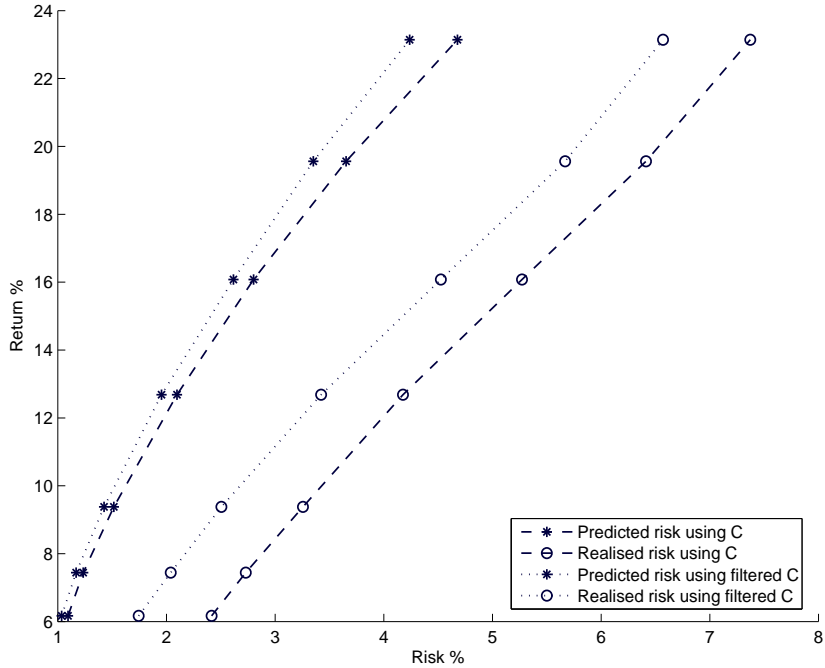}
\caption{Efficient Frontiers using original and cleaned correlation matrices}
\label{Port_Opt_graph}
\end{center}
\end{figure}

It was noted earlier that where the time series available to estimate cross-correlation matrices are of finite length this leads to measurement noise.  This problem is particularly prevalent with hedge fund data, since only monthly returns are available.  As seen in Figure~\ref{Port_Opt_graph} the realised risk is, on average, 292$\%$ of the predicted risk.  This has obvious consequences for risk management.  However, by cleaning the correlation matrix as described in Section~\ref{Port_Opt_Intro}, the realised risk is 190$\%$ of the predicted risk. This huge improvement is brought about by limiting the correlation matrix to the information band prescribed by RMT. 

It can be seen in Figure~\ref{Port_Opt_graph} that, for some return values, the predicted risk (using the filtered correlation matrix) is actually less than that of the original correlation matrix.  This is due to the constraints imposed on the portfolio, in particular the restriction of no `short-selling' (Section~\ref{Port_Opt_Intro}).  

The use of the cleaned correlation matrix leads to a $35\%$ improvement in the difference between the realised risk and the predicted risk for the optimal portfolio.  So we have shown that, in the case of Hedge Fund returns data, the cleaned correlation matrix is a good choice for portfolio optimisation.
 
\clearpage
\section{Conclusions}
We have illustrated that, even with limited data (105 months of returns data for 49 Hedge Funds), useful information can be extracted from a cross correlation matrix constructed from hedge fund returns.  Significant deviations from Random Matrix Theory predictions are observed, with further analysis showing that there is real strategy information contained within the deviating eigenvalues.  Eigenvector analysis revealed distinct \emph{strategy clustering} in the deviating eigenvectors. These included Emerging Markets and Managed Futures in the largest eigenvector, Equity funds in the second, Currency and Fund of Funds in the final two deviating eigenvectors.  The strategy information in the deviating eigenvalues was then used to clean the correlation matrix, by flattening the eigenvalues from the bulk to their average and holding the deviating eigenvalues the same.  A $35\%$ improvement between the risk of the predicted and realised portfolios was found using this filtering technique.

\appendix
\section{Hedge Fund Strategies}
Strategies employed by the managers in the sample considered:
\begin{table}[htbp!]
	\centering
		\begin{tabular}{ll}
\bf Strategies &	\bf Number of Funds \\
Asia excluding Japan Long/Short Equities &	2 \\
Convertible \& Equity Arbitrage	& 2 \\
Currency	& 7 \\
Emerging Markets	& 6 \\
European Long/Short Equity & 	10\\
Fixed Income	& 1 \\
Global Equity&	5\\
Japan Market Neutral	& 1\\
Macro &	3\\
Managed Futures	& 11\\
Self-Invested Fund of Funds	& 1
		\end{tabular}
	\caption{Hedge Fund Strategies}
	\label{tab:HFstrats}
\end{table}

\end{document}